\documentclass[aps,twocolumn,pre,eqsecnum]{revtex4}
\usepackage{graphpap}
\usepackage[dvips]{graphicx}
\usepackage[dvips]{graphics}
\usepackage{color}
\usepackage{hyperref}
\usepackage{amsmath}
\usepackage{ulem}
\usepackage{float}
\usepackage{textcomp}
\usepackage{siunitx}

\graphicspath{{./figures/}}

\usepackage{nomencl}
\makenomenclature

\begin{document}

%--------------------------------------------------
\title{Force sensitivity of multilayer graphene optomechanical devices
}
%--------------------------------------------------

\author{P. Weber, J. G\"uttinger, A. Noury, J. Vergara-Cruz, A. Bachtold\footnotemark[1]}

\footnotetext[1]{Corresponding author. E-mail: adrian.bachtold@icfo.es.}

\affiliation{ICFO-Institut de Ciencies Fotoniques, The Barcelona Institute of Science and Technology, 08860 Castelldefels (Barcelona), Spain}

\date{\today}

\begin{abstract}
Mechanical resonators based on low-dimensional 
materials are promising for force and mass sensing experiments.
The force sensitivity in these ultra-light resonators is often limited
by the imprecision in the measurement of the vibrations, the fluctuations
of the mechanical resonant frequency, and the heating induced by the 
measurement. Here, we strongly couple multilayer graphene resonators 
to superconducting cavities in order to achieve a displacement sensitivity
of $1.3\,$fm\,Hz$^{-1/2}$. This coupling also allows us to damp the
resonator to an average phonon occupation of $7.2$. Our best force
sensitivity, $390\,$zN\,Hz$^{-1/2}$ with a 
bandwidth of 200\,Hz, is achieved by balancing measurement
imprecision, optomechanical damping, and heating. Our results hold 
promise for studying the quantum capacitance of graphene, its 
magnetization, and the electron and nuclear spins of molecules 
adsorbed on its surface.
\end{abstract}

\maketitle
\section{Introduction}
Considerable effort has been devoted to developing mechanical resonators based on low-dimensional materials, such as carbon nanotubes~\cite{sazanova2004,jensen2008,chiu2008,lassagne2009,steele2009a,gouttenoire2010,chaste2012,moser2013,ganzhorn2013,moser2014,benyamini2014,haekkinen2015}, semiconducting nanowires~\cite{ayari2007,gil-santos2010,arcizet2011,nichol2012,nichol2013,sansa2014,gloppe2014,montinaro2014,mathew2015,nigues2015},
graphene~\cite{bunch2007,chen2009,eichler2011a,miao2014,singh2014,song2014,weber2014},
 and monolayer semiconductors~\cite{lee2013, vanleeuwen2014, wang2014}. The specificity of these resonators is their small size and their ultra-low mass, which enables sensing of force and mass with unprecedented sensitivities~\cite{moser2014,chaste2012}. Such high-precision sensing capabilities hold promise for studying physical phenomena in new regimes that have not been explored thus far, for instance, in spin physics~\cite{rugar2004}, quantum electron transport~\cite{bleszynski2009, chen2015}, light-matter interaction~\cite{gloppe2014} and surface science~\cite{wang2010, tavernarakis2014}. However, the transduction of the mechanical vibrations of nanoscale mechanical systems into a measurable electrical or optical output signal is challenging. As a result, force and mass sensing is often limited by the imprecision in the measurement of the vibrations, and cannot reach the fundamental limit imposed by thermo-mechanical noise.

A powerful method to obtain efficient electrical readout of small resonators is to amplify the interaction between mechanical vibrations and the readout field using a superconducting microwave cavity~\cite{song2014,weber2014,singh2014}. Increasing the field in the cavity improves the readout sensitivity and eventually leads to dynamical back-action on the thermo-mechanical noise. This effect has been studied intensively on comparatively large micro-fabricated resonators, resulting for instance in enhanced optomechanical damping~\cite{arcizet2006, gigan2006}, ground-state cooling of mechanical vibrations~\cite{teufel2011b,chan2011}, and displacement imprecision below the standard quantum limit~\cite{teufel2009, anetsberger2010}. Another phenomenon often observed when detecting and manipulating the motion of mechanical resonators is the induced heating that can occur through Joule dissipation and optical adsorption~\cite{song2014, meenehan2014}. Heating is especially prominent in tiny mechanical resonators because of their small heat capacity. An additional difficulty in characterizing mechanical vibrations is related to the fluctuations of the mechanical resonant frequency, also called frequency noise, which are particularly sizable in small resonators endowed with high quality factors $Q$~\cite{moser2014}.

\begin{figure}
	\includegraphics[width=8.5cm]{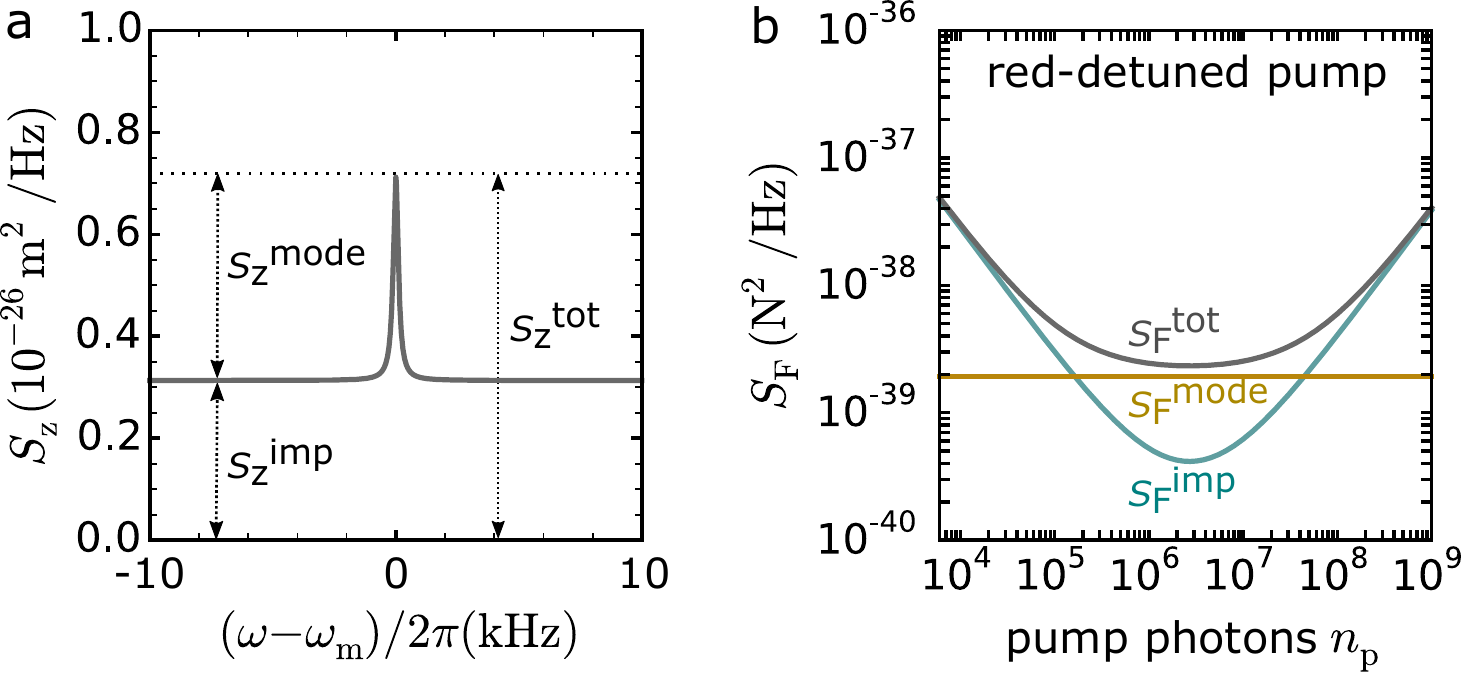}
	\caption{\textbf{Mechanical displacement and force sensitivity.} (\textbf{a}) Mechanical displacement spectrum $S_\mathrm z$ close to the mechanical resonance frequency $\omega_\mathrm m/2\pi$. The total displacement spectral density $S_\mathrm z^\mathrm{tot}$ at $\omega_\mathrm m$ is the sum of the displacement noise $S_\mathrm z^\mathrm{mode}(\omega_\mathrm m)$ and the displacement imprecision $S_\mathrm z^\mathrm{imp}$. (\textbf{b}) Corresponding force sensitivity $S_\mathrm F^\mathrm{tot} = S_\mathrm F^\mathrm{mode}+S_\mathrm F^\mathrm{imp}$ (dark grey). The individual components are the thermal force noise $S_\mathrm F^\mathrm{mode}$ (dark yellow) and the imprecision force noise $S_\mathrm F^\mathrm{imp}$ (turquois), given by Eqs.~\ref{eq:disp_force} and \ref{eq:imp_force}, respectively. The quantum back-action noise is neglected for simplicity. For the plots most of the parameters are those of device B, but we estimate the mass assuming that the graphene flake is a single layer. Further we choose $n_\mathrm{add}=0.5$, $T_\mathrm{bath}=0.015\,$K, and $n_\mathrm p = 2\cdot 10^5$ in \textbf a (see text).}
	\label{fig1}
\end{figure}

Here we study the force sensitivity of multilayer graphene mechanical resonators coupled to superconducting cavities. In particular, we quantify how the force sensitivity is affected by dynamical back-action, Joule heating, and frequency noise upon increasing the number of pump photons inside the cavity. We demonstrate a force sensitivity of $(S_\mathrm F^\mathrm{tot})^{1/2}=390 \pm 30 \,$zN\,Hz$^{-1/2}$, of which $\approx 50\%$ arises from thermo-mechanical noise and $\approx 50\%$ from measurement imprecision. The force sensitivity tends to be limited by measurement imprecision and frequency noise at low pump power, and by optomechanical damping and Joule heating at high pump power.

\section{Results}
\subsection{Thermal force noise and imprecision force noise}
A fundamental limit of force sensing is set by the thermo-mechanical noise of the eigenmode that is measured. According to the fluctuation-dissipation theorem, the associated thermal force noise is white and is quantified by\\ 
\begin{equation}
	\label{eq:disp_force}
	S_\mathrm F^\mathrm{mode}= 4k_\mathrm B T_\mathrm{mode} m_\mathrm{eff} \mathit{\Gamma}_\mathrm{eff}^\mathrm{spectral}
\end{equation}\\
where $T_\mathrm{mode}$ is the temperature of the mechanical eigenmode, and $ m_\mathrm{eff}$ is its effective mass ~\cite{mamin2001,moser2013}. This force noise is transduced into a mechanical resonance with line width $\mathit{\Gamma}_\mathrm{eff}^\mathrm{spectral}$ and height $S_\mathrm z^\mathrm{mode}$ in the displacement spectrum (Fig.~\ref{fig1}). Importantly, Eq.~\ref{eq:disp_force} shows that the low mass of graphene decreases the size of the thermo-mechanical force noise. However, a drawback of tiny resonators with high $Q$-factors is their tendency to feature sizable frequency noise that broadens the resonance and, therefore, increases the size of the force noise \cite{moser2014,zhang2014}.

\begin{figure}
	\includegraphics[width=8.5cm]{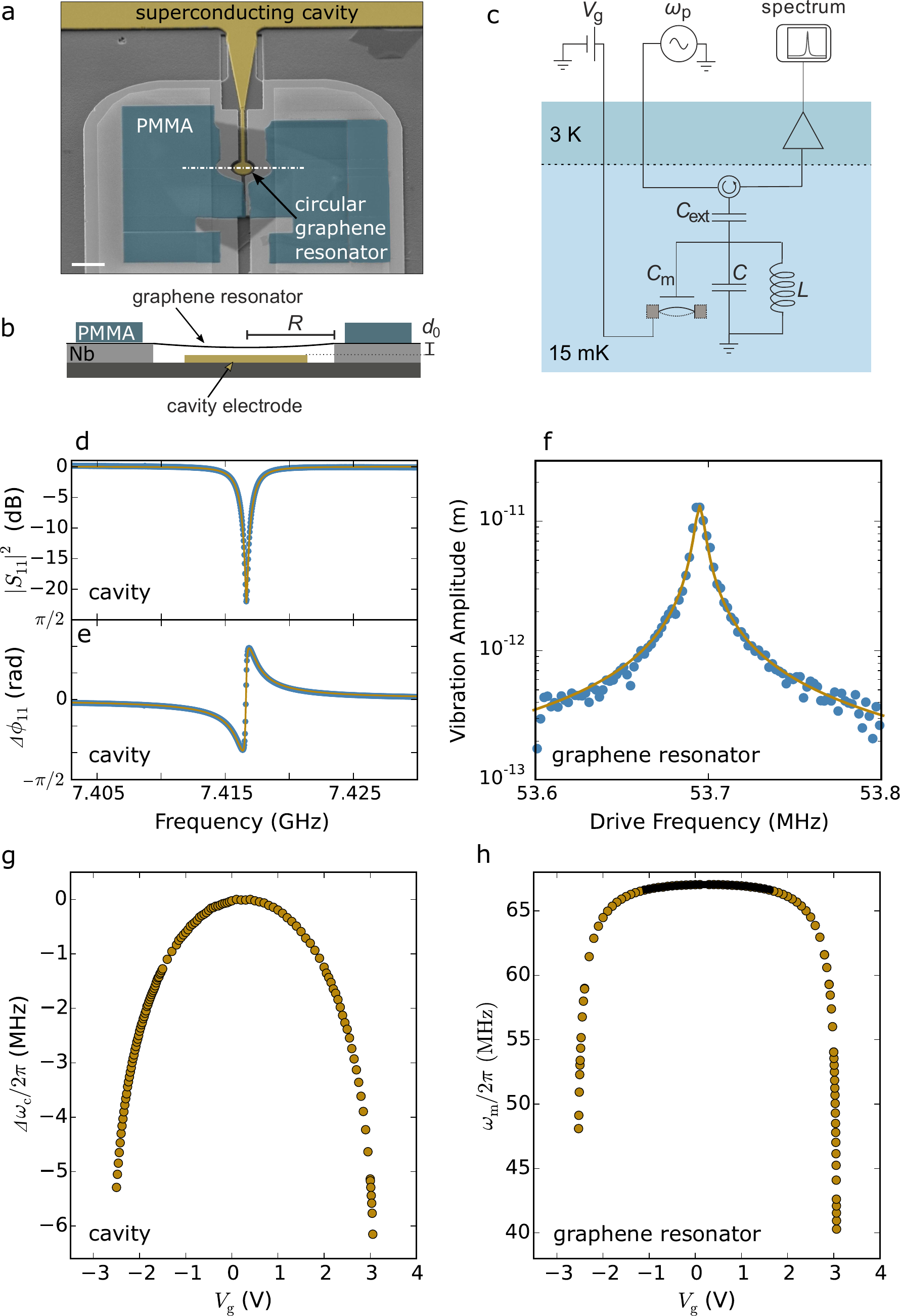}
	\caption{\textbf{Device and characterization.} (\textbf{a}) False-color image of the device.  The cavity is coloured in dark yellow. The graphene flake is clamped in between niobium support electrodes (grey) and cross-linked PMMA (turquois). The scale bar is $5\,\mu$m. (\textbf{b}) 
	Schematic cross-section of the graphene resonator along the white dashed dotted line in \textbf a. (\textbf{c}) Schematic of the measurement circuit. The graphene mechanical resonator is coupled to the superconducting LC cavity through the capacitance $C_\mathrm m$. The separation $d$ between the suspended graphene flake and the cavity counter electrode is controlled by the constant voltage $V_\mathrm g$. The cavity is pumped with a pump tone at $\omega_\mathrm p$ and the output signal is amplified at $3\,$K. (\textbf{d}) Reflection coefficient $|S_{11}|^2$ and (\textbf{e}) reflected phase $\mathit{\Delta} \phi_{11}$ of the superconducting cavity of device A at $V_\mathrm g = 3.002\,$V. The dark yellow lines are fits to the data using $\kappa_\mathrm{int}/2\pi = 950\,$kHz and $\kappa_\mathrm{ext}/2\pi = 850\,$kHz using Eq.~\ref{eq:cavity} (see Methods). (\textbf{f}) Driven vibration amplitude of the graphene resonator of device A as a function of drive frequency. The driving voltage is $22\,$nV and $V_\mathrm g = 3.002\,$V. The dark yellow line is a lorentzian fit to the data. (\textbf{g}) Resonant frequency $\omega_\mathrm c/2\pi$ of the superconducting cavity as a function of $V_\mathrm g$. (\textbf{h}) Resonant frequency $\omega_\mathrm m/2\pi$ of the graphene resonator as a function of $V_\mathrm g$. The black line is the $V_\mathrm g$ dependence of $\omega_\mathrm m$ expected from electrostatic softening (see Supplementary Note 1).}
	\label{fig2}
\end{figure}

Measuring mechanical vibrations with high accuracy is key to resolving small forces, since the imprecision in the measurement contributes to the force sensitivity. The force sensitivity $S_\mathrm F^\mathrm{tot}$ is given by the sum of the thermal force noise $S_\mathrm F^\mathrm{mode}$ and the imprecision force noise $S_\mathrm{F}^\mathrm{imp}$, where the latter is the result of the white noise background with strength $S_\mathrm z^\mathrm{imp}$ in the displacement spectrum (Fig.~\ref{fig1}a). The challenge with mechanical resonators based on low-dimensional systems is to reach the limit $S_\mathrm{F}^\mathrm{imp} < S_\mathrm F^\mathrm{mode}$. When detecting the motion of graphene resonators with microwave cavities, one typically operates in the resolved sideband limit~\cite{weber2014,song2014,singh2014}, where the cavity decay rate $\kappa$ is significantly smaller than the mechanical resonance frequency $\omega_\mathrm m$. This is interesting for force sensing, because pumping on the red sideband allows to  enhance the mechanical damping rate by $\mathit{\Gamma}_\mathrm{opt}$, and therefore to reduce the harmful effect of frequency noise, as we will discuss below. In addition, this allows to increase the measurement bandwidth, as is often done in magnetic resonance force microscopy experiments~\cite{rugar2004} while keeping $S_\mathrm F^\mathrm{mode}$ constant. The drawback of red sideband pumping compared to pumping at the cavity resonant frequency is an increased imprecision force noise at high pump powers. In the red-detuned pump regime, the measurement imprecision contributes to the force sensitivity by the amount\\
\begin{equation}
	\label{eq:imp_force}
	S_\mathrm F^\mathrm{imp} = \hbar \omega_\mathrm m m_\mathrm{eff} \frac{\kappa}
	{\kappa_\mathrm{ext}} \frac{\left(\mathit{\Gamma}_\mathrm{m}^\mathrm{spectral} + 4 n_\mathrm p g_0^2/\kappa \right)^2}{4 n_\mathrm p g_0^2/\kappa} \left(n_\mathrm{add} + \frac{1}{2} \right),  
\end{equation}\\
with $\kappa_\mathrm{ext}$ the external coupling rate of the cavity, $n_\mathrm{add}$ the noise added by the amplifier chain at the output of the device, $\mathit{\Gamma}_\mathrm{m}^\mathrm{spectral}$ the intrinsic line width of the resonator, $n_\mathrm p$ the number of pump photons in the cavity, and $g_0$ the single-photon optomechanical coupling. Figure~\ref{fig1}b shows the pump power dependence of the force sensitivity $S_\mathrm F^\mathrm{tot}$ expected in the absence of Joule heating and frequency noise. The increase of $S_\mathrm F^\mathrm{tot}$ at high $n_\mathrm p$ is due to the dynamical back-action, which enhances the mechanical line width by $\mathit{\Gamma}_\mathrm{opt} = 4 n_\mathrm p g_0^2/\kappa$.

\subsection{Device characterization}
Our devices consist of a suspended graphene mechanical resonator capacitively coupled to a superconducting niobium cavity (Fig.~\ref{fig2}a-c). The graphene resonators are circular with a radius of $R \approx \SI{1.6}{\micro\meter}$. Here we present data of 2 devices. The graphene resonator of device A has a thickness of approximately 25~layers, and the one of device B 5-6~layers. This corresponds respectively to an effective mass of $m_\mathrm{eff} = (4.1 \pm 0.8)\cdot 10^{-17}\,$kg and $(9.6 \pm 0.8) \cdot 10^{-18}\,$kg. The uncertainty results from extracting the mass with different methods including optical contrast measurements, thickness measurements with atomic force microscopy (AFM) and the measured electrostatic softening of the mechanical resonators (see Supplementary Note 1 and Supplementary Equation 2). The fundamental mode of devices A and B vibrates
at $\omega_\mathrm m/2\pi = 67\,$MHz and $\omega_\mathrm m/2\pi = 46\,$MHz at $V_\mathrm g=0\,$V, respectively. Here $V_\mathrm g$ is the constant voltage applied between the graphene flake and the superconducting cavity. In order to improve the attachment of the graphene flake to its support, we clamp it between cross-linked poly(methyl metracylate)(PMMA) and the contact electrodes; the detailed fabrication is described elsewhere \cite{weber2014}. The separation between the graphene resonator and the cavity counter electrode at $V_\mathrm g=0\,$V is assumed to be equal to the hole depth, which is typically $d_0 \approx 85\,$nm in our devices as measured with AFM. Varying $V_\mathrm g$ allows us to tune the separation between the graphene resonator and the cavity counter electrode  \cite{chen2009, singh2010, bao2012, chen2013, weber2014}, modifying the graphene-cavity capacitance, the cavity frequency $\omega_\mathrm c$, and $\omega_\mathrm m$ (Figs.~\ref{fig2}g,h). The superconducting cavity is a coplanar waveguide resonating at about $\omega_\mathrm c/2\pi = 7.4\,$GHz. We choose a single-port, quarter wavelength, reflection geometry, so that the cavity is connected to ground on one end, allowing to apply a well defined constant voltage between the cavity and the graphene flake. The other end of the cavity is coupled to a transmission line via a capacitor $C_\mathrm{ext}$ with a coupling rate $\kappa_\mathrm{ext} = 2\pi \times 850\,$kHz for device A; the total cavity decay rate is $\kappa = \kappa_\mathrm{ext} + \kappa_\mathrm{int} =  2\pi \times 1.8\,$MHz (see Methods). Here $\kappa_\mathrm{int}$ accounts for the internal energy loss.

We detect the vibrations of the graphene resonator with high precision by pumping the cavity with an electromagnetic field, and probing its mechanical sideband. This sideband is generated by the capacitive modulation of the pump field at frequency $\omega_\mathrm p/2\pi$ by the graphene vibrations at $\omega_\mathrm m/2\pi$. We usually set $\omega_\mathrm p=\omega_\mathrm c - \omega_\mathrm m$
and probe the electromagnetic field that exits the cavity at $\omega_\mathrm c$. We measure the device at the cryostat base temperature of $15\,$mK if not stated otherwise.
The cavity output field is amplified with a high electron-mobility-transistor
(HEMT) mounted at the 3~K stage of the cryostat. Mechanical noise
spectra are detected with a spectrum analyzer at room temperature. For a detailed description of the measurement setup see Supplementary Fig.~1 and Supplementary Note 2.
In addition, we perform ring-down measurements to determine the
mechanical dissipation rate $\mathit{\Gamma}_\mathrm{eff}^\mathrm{decay}$ of the graphene resonator.
Spectral measurements are not suitable for quantifying reliably $\mathit{\Gamma}_\mathrm{eff}^\mathrm{decay}$
because of the potentially substantial frequency noise of graphene
resonators. 

\begin{figure}
	\includegraphics[width=8.5cm]{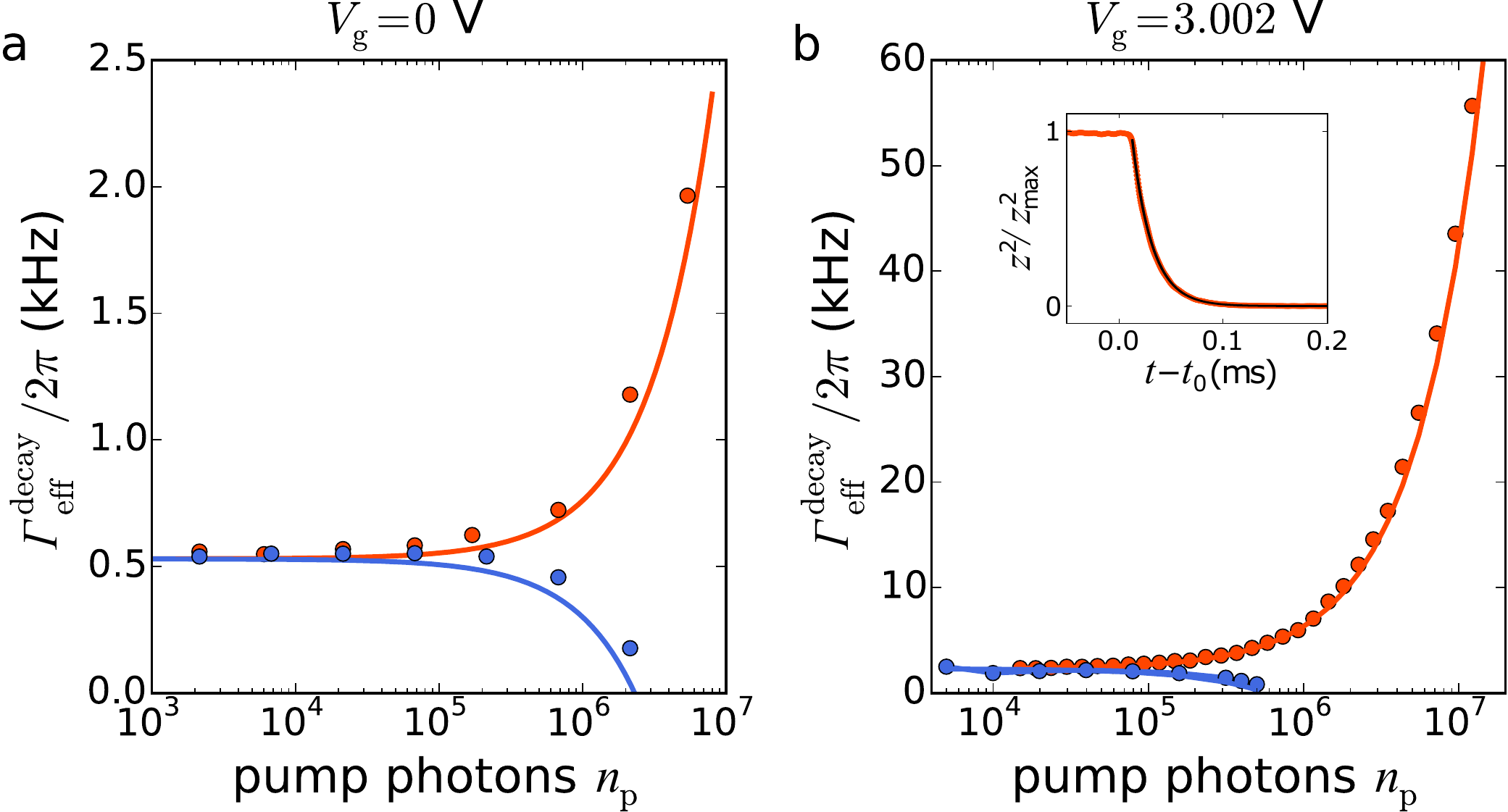}
	\caption{\textbf{Effective mechanical energy decay rates extracted from ring-down measurements.} Mechanical dissipation rate $\mathit{\Gamma}_\mathrm{eff}^\mathrm{decay}$ measured on device A with the ring-down technique as a function of the number $n_\mathrm p$ of pump photons in the cavity at $V_\mathrm g=0\,$V and $V_\mathrm g=3.002\,$V where $n_\mathrm p$ is proportional to the microwave power $P_\mathrm{in}$ applied at the input of the cryostat (see Supplementary Note 3). Red and blue data points correspond to red and blue detuned pumping, respectively. The measurements are well described by $\mathit{\Gamma}_\mathrm{eff}^\mathrm{decay}=\mathit{\Gamma}_\mathrm{m}^\mathrm{decay} \pm \mathit{\Gamma}_\mathrm{opt}$ (red and blue lines) using $g_0/2\pi=9.7\,$Hz in \textbf{a} and $g_0/2\pi=42.6\,$Hz in \textbf{b}. The inset in \textbf{b} shows a ring-down measurement for $n_\mathrm p = 1.4 \cdot 10^6$. We plot the normalized vibration amplitude as a function of time $t$. The resonator is driven with a capacitive driving force for $t<t_0$. At $t_0$ the drive is switched off and the vibration amplitude decays freely ($t>t_0$). We fit the data with an exponential decay (black line) using $z^2(t) = z^2_\mathrm{max} exp(-\frac{t-t_0}{\tau})$ with a decay rate $\mathit{\Gamma}_\mathrm{eff}^\mathrm{decay} = 1/\tau = 2\pi \cdot 8.4\,$kHz. The vibration amplitude in ring-down measurements is larger than that in undriven displacement spectra, so that the motion in ring-down measurements can be resolved with lower $n_\mathrm p$. 
	}
	\label{fig3}
\end{figure}

We characterize the single-photon optomechanical coupling and show that the coupling can be significantly enhanced by deflecting the membrane towards the cavity electrode. For this, we quantify the optomechanical
scattering rate $\mathit{\Gamma}_\mathrm{opt}$ using ring-down measurements
at $V_\mathrm g= 0\,$V and $V_\mathrm g=3.002\,$V for device A.
Figures~\ref{fig3}a,b show the measured dissipation rate
$\mathit{\Gamma}_\mathrm{eff}^\mathrm{decay}$ as a function of cavity pump photon number
$n_\mathrm p$ for blue and red detuned pumping. The measurements
are well described by $\mathit{\Gamma}_\mathrm{eff}^\mathrm{decay} = \mathit{\Gamma}_\mathrm{m}^\mathrm{decay} \pm
\mathit{\Gamma}_\mathrm{opt}$ where $\mathit{\Gamma}_\mathrm{m}^\mathrm{decay}$ corresponds to the intrinsic mechanical dissipation rate, and $\pm$ to red and blue detuned
pumping at $\omega_\mathrm p=\omega_\mathrm c \mp \omega_\mathrm m$,
respectively.
By increasing $V_\mathrm g$ from $0$ to $3.002\,$V we obtain a strong
increase of the optomechanical coupling from $g_0=2\pi \times 9.7\,$Hz
to $g_0=2\pi \times 42.6\,$Hz. We estimate that the separation $d$ between the membrane and the cavity counter electrode is reduced from $88\,$nm to $33\,$nm when varying $V_\mathrm g$ from $0$ to $3.002\,$V. The calibration of both $g_0$ and
$n_\mathrm p$  is robust, while the quantification of the reduction of $d$ is approximative; see Supplementary Notes 1 and 3, Supplementary Fig.~2 and Supplementary Equations 1, 3-5.

\subsection{Thermal calibration and sideband cooling}
\begin{figure}
	\includegraphics[width=8.5cm]{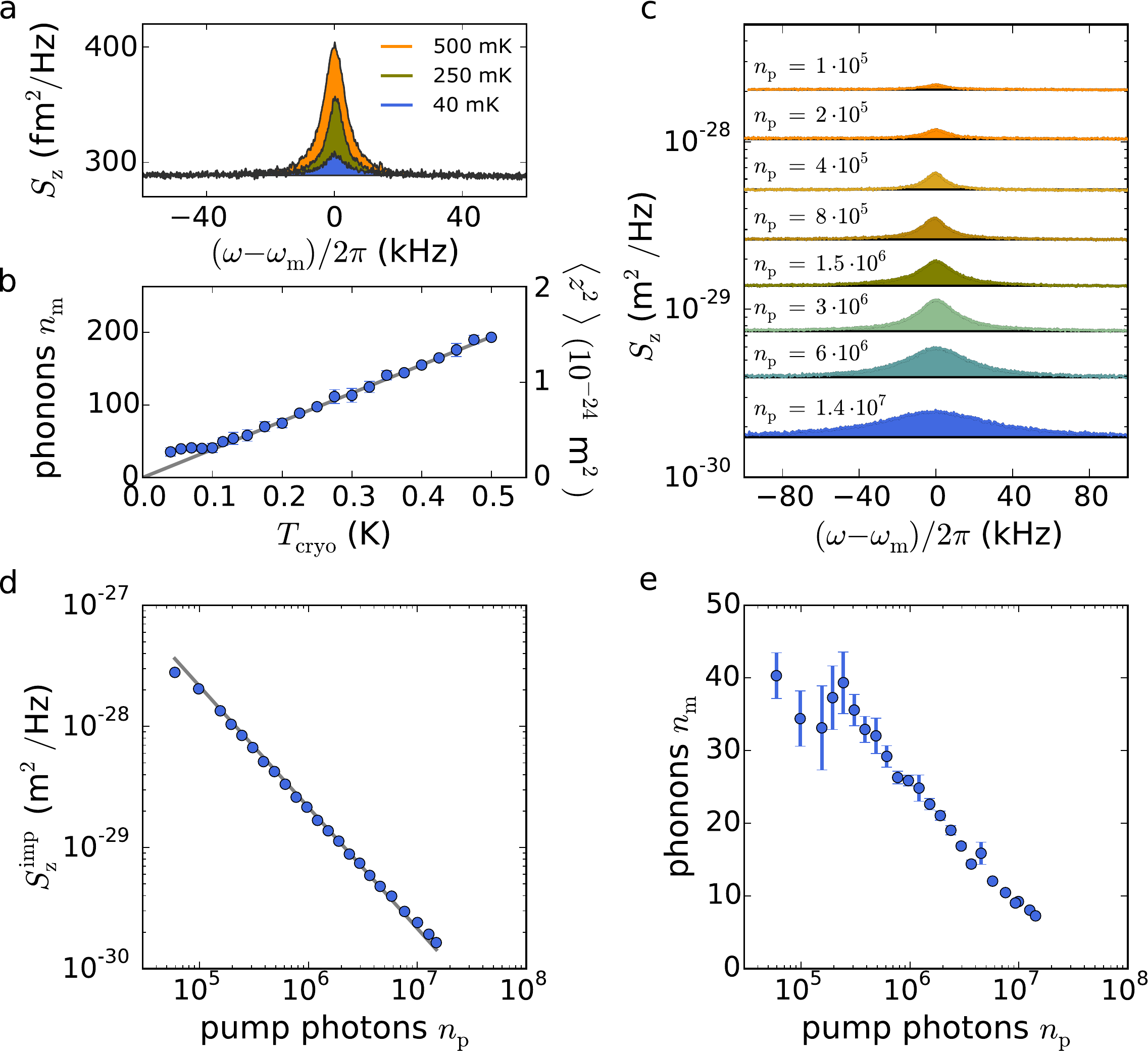}
	\caption{\textbf{Thermal calibration and sideband cooling of fundamental mechanical mode with red-detuned pumping.} (\textbf{a}) Selected thermo-mechanical noise spectra for different temperatures and $n_\mathrm p = 6\cdot 10^4$. (\textbf{b}) Plot of the measured mechanical mode temperature of device A, expressed in phonon occupation $n_\mathrm m$, as a function of cryostat temperature at $V_\mathrm g =3.002\,$V where $\omega_\mathrm m/2\pi=53.7\,$MHz and $n_\mathrm p = 6\cdot 10^4$. On the right y-axis, we display the variance of the vibration amplitude $\left\langle z^2 \right\rangle$, which is obtained by integrating the thermal resonance, as is shown in \textbf a. The phonon occupation is quantified with $\left\langle z^2 \right\rangle = \frac{\hbar}{m_\mathrm{eff} \omega_\mathrm m} n_\mathrm m$ (see Supplementary Note 3). The error bars are given by the standard deviation of 5 spectral measurements. (\textbf{c}) Mechanical displacement spectral density $S_\mathrm{z}$ measured for different pump photon number. The cryostat temperature is $15\,$mK. Note that the curves are not offset.		
		(\textbf{d}) Displacement imprecision as a function of cavity pump photon population. The line is a fit of Eq.~\ref{eq:disp_imp} with $n_\mathrm{add} = 32$. (\textbf{e}) Average phonon number $n_\mathrm m$ as a function of $n_\mathrm p$. The error bars are given by the standard deviation of 5 spectral measurements.} 
	\label{fig4}
\end{figure}

In order to calibrate the mechanical phonon occupation and the mode temperature $T_\mathrm {mode}$, we measure the mechanical thermal motion
spectrum while varying the cryostat temperature \cite{teufel2011b}.
This is done by pumping the cavity with a weak pump tone on the red sideband.
The integrated area of the thermal resonance is proportional to the mode temperature according to the equipartition theorem.
For temperatures above $100\,$mK the area is linearly proportional to the cryostat temperature, showing that the mode is in thermal equilibrium with the cryostat (Fig.~\ref{fig4}b).
This linear dependence serves as a precise calibration to
relate the resonance area to the averaged phonon occupation $n_\mathrm m$ and the mode temperature $T_\mathrm {mode}$. Below
$100\,$mK the mechanical mode does not thermalize well with the
cryostat. The origin of this poor thermalization at low temperature may be related to the heating induced by the pump field (see below)~\cite{song2014}, and a non-thermal force noise~\cite{rocheleau2010} such as the electrostatic force noise related to the voltage noise in the device.
As a next characterization step, we investigate the mechanical phonon occupation when increasing the power of the pump tone on the red
sideband and keeping the temperature of the cryostat constant at $T_\mathrm{cryo}=15\,$mK. The measured resonance gets broader and its area smaller (Fig.~\ref{fig4}c), showing that the mechanical mode
is damped and cooled \cite{arcizet2006, gigan2006}. At the largest available pump power, the phonon occupation reaches $n_\mathrm m=7.2 \pm
0.2$ (Fig.~\ref{fig4}e). This is the lowest phonon occupation reached in a mechanical resonator based on graphene ~\cite{song2014, barton2012, singh2014}.
The error in the estimation of $n_\mathrm{m}$ is given by the standard error obtained from 5 successive spectral measurements.

\subsection{Displacement sensitivity and force sensitivity}
The improved coupling allows us to achieve also an excellent
displacement sensitivity $S_\mathrm z^\mathrm{imp}$ (Fig.~\ref{fig4}d). At the largest pump power, we obtain $(S_\mathrm
	z^\mathrm{imp})^{1/2}=1.3 \pm 0.2\,$fm\,Hz$^{-1/2}$, which compares favorably to previous works \cite{barton2012, singh2014, cole2015}.  The
error in $S_\mathrm{z}^\mathrm{imp}$ is given by the uncertainty in the
estimation of $m_\mathrm{eff}$. We obtain $S_\mathrm z^\mathrm{imp}$ from the
noise floor of the measured power spectral density $S_\mathrm N$
using $S_\mathrm z^\mathrm{imp} = \frac{S_\mathrm N}{\hbar \omega_\mathrm c} \frac{\kappa^2}{2\kappa_\mathrm{ext}} \frac{z_\mathrm{zp}^2}{g_0^2} \frac{1}{n_\mathrm p}$ with $z_\mathrm{zp} = \sqrt{\hbar/2m_\mathrm{eff}\omega_\mathrm m}$ the zero-point motion amplitude \cite{singh2014}. The displacement sensitivity scales as $1/n_\mathrm p$ (Fig.~\ref{fig4}d) . By comparing the measurement to the expected displacement sensitivity\\ 
\begin{equation}
	S_\mathrm z^\mathrm{imp} = \left( n_\mathrm{add} + \frac{1}{2} \right) \frac{\kappa^2}{2\kappa_\mathrm{ext}} \frac{z_\mathrm{zp}^2}{g_0^2} \frac{1}{n_\mathrm p},
	\label{eq:disp_imp}
\end{equation}\\
we obtain that the equivalent noise added by the amplifier chain is $n_\mathrm{add} = 32$. This is a reasonable value for a HEMT amplifier mounted at 3~K~\cite{castellanos2008, teufel2009}.

\begin{figure*}
	\includegraphics[width=17cm]{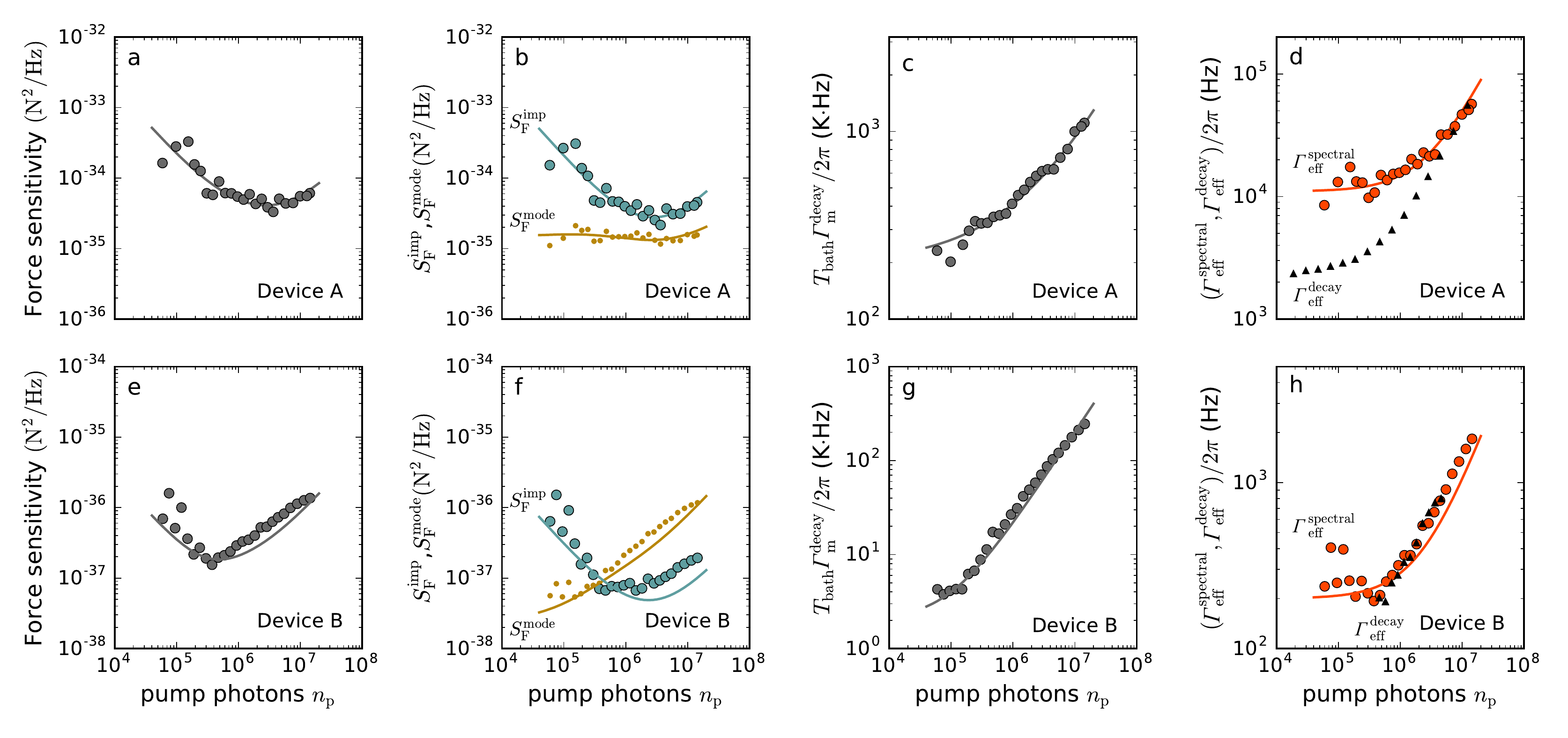}
	\caption{\textbf{Characterization of the imprecision force noise, the thermal force noise and the total force sensitivity.} (\textbf{a}) Force sensitivity $S_\mathrm F^\mathrm{tot}=S_\mathrm F^\mathrm{imp} + S_\mathrm F^\mathrm{mode}$ as a function of cavity pump photon population measured when pumping the cavity on the red sideband. (\textbf{b}) Imprecision force noise $S_\mathrm F^\mathrm{imp}$ (turquois) and thermal force noise $S_\mathrm F^\mathrm{mode}$ (dark yellow) versus $n_\mathrm p$. The data in \textbf{a,b} are fitted to Eqs.~\ref{eq:imp_force}, \ref{eq:freq_noise}. (\textbf{c}) Product of the bath temperature $T_\mathrm{bath}$ and the intrinsic mechanical decay rate $\mathit{\Gamma}_\mathrm m^\mathrm{decay}$ as a function of cavity pump photon occupation. The line is a fit to the data. (\textbf{d}) Effective spectral mechanical line width $\mathit{\Gamma}_\mathrm{eff}^\mathrm{spectral}$ and energy decay $\mathit{\Gamma}_\mathrm{eff}^\mathrm{decay}$ as a function of $n_\mathrm p$. The data are fitted to $\mathit{\Gamma}_\mathrm{eff}^\mathrm{spectral} = \mathit{\Gamma}_\mathrm{eff}^\mathrm{decay} + \delta \mathit{\Gamma}_\mathrm{noise}$ with $\delta \mathit{\Gamma}_\mathrm{noise}/2\pi = 8.7\,$kHz (red line). (\textbf{e-h}) Equivalent to \textbf{(a-d)} but for device B. The lowest value for the force sensitivity in \textbf e is $(S_\mathrm F^\mathrm{tot})^{1/2}=390 \pm 30\,$zN\,Hz$^{-1/2}$. In \textbf e and \textbf f the data are fitted with $n_\mathrm{add} = 22$ and in \textbf h we use $g_0/2\pi = 7.3\,$Hz, $\kappa/2\pi = 2.5\,$MHz and $\delta \mathit{\Gamma}_\mathrm{noise}/2\pi = 0.145\,$kHz. All the measurements on device A are performed at $V_\mathrm g = 3.002\,$V and on device B at $V_\mathrm g = 0\,$V. The cryostat temperature is $15\,$mK.}  
	\label{fig5}
\end{figure*}

We now quantify the force sensitivity as a function of the microwave pump power (Figs.~\ref{fig5}a,e).  Since the mechanical resonances in the measured displacement spectra are well described by Lorentzian line shapes, the thermal force noise is quantified using $S_\mathrm F^\mathrm{mode} = S_\mathrm z^\mathrm{mode}(\omega_\mathrm m) /|\chi(\omega_\mathrm m)|^2$ with the effective mechanical susceptibility $|\chi(\omega_\mathrm m)|^2=1/(m_\mathrm{eff} \omega_\mathrm m \mathit{\Gamma}_\mathrm{eff}^\mathrm{spectral})^2$. Similarly, we obtain the imprecision force noise with $S_\mathrm F^\mathrm{imp} = S_\mathrm z^\mathrm{imp} /|\chi(\omega_\mathrm m)|^2$. The best force sensitivity we achieve for device A is $(S_\mathrm F^\mathrm{tot})^{1/2}=5.8\,$aN\,Hz$^{-1/2}$  with a mechanical bandwidth of $20\,$kHz (Fig.~\ref{fig5}a,d). In device B we reach a force sensitivity of $(S_\mathrm{F}^\mathrm{tot})^{1/2}=390 \pm 30 \,$zN\,Hz$^{-1/2}$ with a mechanical bandwidth of $0.2\,$kHz (see Figs.~\ref{fig5}e,h). The error in the estimation of the force sensitivity is obtained from both the uncertainty in the mass and the fluctuations in the measurement of 
$S_\mathrm F^\mathrm{tot}$, which we evaluate by calculating the standard error of 10 measurements. This force sensitivity
compares favorably with the best sensitivities obtained with micro-fabricated resonators ($(S_\mathrm{F}^\mathrm{tot})^{1/2}=510\,$zN\,Hz$^{-1/2}$) \cite{mamin2001,teufel2009}, albeit it is not as good as
that of resonators based on carbon nanotubes \cite{moser2013, moser2014}. Compared to previous devices, the mechanical bandwidth of graphene resonators is much higher, which enables faster detection of sudden force changes.

\section{Discussion}
We plot both $S_\mathrm F^\mathrm{mode}$ and $S_\mathrm{F}^\mathrm{imp}$ as a function of cavity pump photon population in Fig.~\ref{fig5}b. As expected, the imprecision force noise decreases at low $n_\mathrm p$ and increases at high $n_\mathrm p$ due to the enhanced damping caused by the optomechanical back-action. The thermal force noise $S_\mathrm F^\mathrm{mode}$ appears roughly constant when varying $n_\mathrm p$ as a result of the competing effects of Joule heating and frequency noise. 
Joule heating is caused by the microwave current in the graphene flake induced by the pump field. This results in the increase of the temperature $T_\mathrm{bath}$ of the thermal bath coupled to the mechanical mode as well as the mechanical dissipation rate~\cite{song2014, miao2014}. We can infer the product  $ T_\mathrm{bath} \cdot \mathit{\Gamma}_\mathrm m^\mathrm{decay} $ from the measurements of $n_\mathrm m$ and 
$\mathit{\Gamma}_\mathrm{eff}^\mathrm{decay}$ in Figs.~\ref{fig3}b, \ref{fig4}e using 
\begin{equation}
	T_\mathrm{bath} \mathit{\Gamma}_\mathrm m^\mathrm{decay} = T_\mathrm{mode} \mathit{\Gamma}_\mathrm{eff}^\mathrm{decay} =  n_\mathrm m \mathit{\Gamma}_\mathrm{eff}^\mathrm{decay} \cdot \frac{\hbar \omega_\mathrm m}{k_\mathrm B}.
	\label{eq:T_mode}
\end{equation}\\
When increasing the pump power, Joule heating significantly increases the product $T_\mathrm{bath} \mathit{\Gamma}_\mathrm m^\mathrm{decay}$ (Fig.~\ref{fig5}c), and therefore the size of the thermal force noise (Eq.~\ref{eq:disp_force}). We see next that the effect of frequency noise leads to the opposite dependence of the thermal force noise on pump power.  
Frequency noise enhances the spectral line width by the amount $\delta \mathit{\Gamma}_\mathrm{noise}$,   
\begin{equation}
	\label{eq:noise}
	\mathit{\Gamma}_\mathrm{eff}^\mathrm{spectral} = \mathit{\Gamma}_\mathrm{eff}^\mathrm{decay}+\delta \mathit{\Gamma}_\mathrm{noise},
\end{equation}\\
when the fluctuations of the resonant frequency are described by a white noise~\cite{moser2013}.
The measurements of $\mathit{\Gamma}_\mathrm{eff}^\mathrm{spectral}$ and $\mathit{\Gamma}_\mathrm{eff}^\mathrm{decay}$ as a function of pump power can be well described by Eq.~\ref{eq:noise} with $\delta \mathit{\Gamma}_\mathrm{noise}/2\pi=8.7\,$kHz (Fig.~\ref{fig5}d). Importantly, Fig.~\ref{fig5}d shows that  $\mathit{\Gamma}_\mathrm{eff}^\mathrm{spectral}$ is comparable to $\mathit{\Gamma}_\mathrm{eff}^\mathrm{decay}$ at large pump power, showing that the relative contribution of $\delta \mathit{\Gamma}_\mathrm{noise}$ to $\mathit{\Gamma}_\mathrm{eff}^\mathrm{spectral}$ gets negligible upon increasing $n_\mathrm p$. 
As the cooling efficiency described by Eq.~\ref{eq:T_mode} remains unaltered by frequency noise (see chapter 7 in~\cite{dykman2012}), the thermal force noise is quantified by
	\begin{equation}
		S_\mathrm F^\mathrm{mode} = 4 k_\mathrm B m_\mathrm{eff} T_\mathrm{bath} \mathit{\Gamma}_\mathrm m^\mathrm{decay} \frac{\mathit{\Gamma}_\mathrm{eff}^\mathrm{spectral}}{\mathit{\Gamma}_\mathrm{eff}^\mathrm{decay}}.
		\label{eq:freq_noise}
	\end{equation}\\
Taking into account the measured effects of Joule heating and frequency noise in Eq.~\ref{eq:freq_noise}, the thermal force noise $S_\mathrm F^\mathrm{mode}$ is expected to remain roughly constant as a function of $n_\mathrm{p}$ (dark yellow line in Fig.~\ref{fig5}b), in agreement with the measurements. 
Overall, the best force sensitivity we achieve in this device is $(S_\mathrm F^\mathrm{tot})^{1/2}=5.8\,$aN\,Hz$^{-1/2}$ at $n_\mathrm p \approx 4 \cdot 10^6$ (Fig.~\ref{fig5}a). While the force sensitivity in this device is primarily limited by the measurement imprecision, the thermal force noise is affected to a large extent by frequency noise at low $n_\mathrm p$ and by Joule heating at high $n_\mathrm p$. 

In device B, the graphene resonator has a lower mass and a narrower mechanical line width, two assets for high force sensitivity (Figs.~\ref{fig5}e-h). The spectral line width corresponds to a mechanical quality factor of $Q \approx 200,000$. In this device we reach a force sensitivity of $(S_\mathrm{F}^\mathrm{tot})^{1/2}=390 \pm 30 \,$zN\,Hz$^{-1/2}$ at $n_\mathrm p \approx 4\cdot 10^5$ (see Figs.~\ref{fig5}e). In an attempt to improve the thermal anchoring of device B compared to device A, the graphene contact electrodes contain an additional Au layer between the graphene and the Nb layer~\cite{fong2013, song2014}. The normal metal layer is expected to increase the thermal conductance between the graphene flake and the contact electrodes through electron diffusion, which allows for better heat dissipation into the contacts. However, Device B is still strongly affected by Joule heating, which substantially increases the value of $S_\mathrm F^\mathrm{mode}$ when increasing the pump power (Figs.~\ref{fig5}f,g).
The heating is so strong that we are not able to reduce the phonon occupation $n_\mathrm m$ with sideband cooling. We attribute the strong heating to the fact that the resonator is significantly thinner than the one of device A and therefore has a smaller heat capacity. The effect of frequency noise on the spectral line width is negligible for pump powers above $n_\mathrm p \approx 4 \cdot 10^5$. We do not know the origin of the frequency noise but it might be related to charged two-level fluctuators in the device. The force sensitivity is here primarily limited by the measurement imprecision at low $n_\mathrm p$, and by the thermo-mechanical force noise and Joule heating at high $n_\mathrm p$.

In the future, the force sensitivity of graphene optomechanical devices can be further improved using a quantum-limited Josephson parametric amplifier~\cite{castellanos2008}. This readout will improve the measurement imprecision, by lowering $n_\mathrm{add}$ in $S_\mathrm{F}^\mathrm{imp}$. In addition, it will be possible to resolve the thermal vibrations with lower pump power, which is crucial to reduce Joule heating while working with low-mass graphene resonators. 
A quantum-limited amplifier with $n_\mathrm{add} = 0.5$ may allow to achieve $47\,$zN\,Hz$^{-1/2}$ force sensitivity at 15\,mK taking the mass of a single-layer graphene resonator with the diameter and the quality factor of device B (Fig.~\ref{fig1}b). With only modest device improvements, it may be possible to probe the fundamental limit of continuous displacement detection imposed by quantum mechanics, since the force noise associated to quantum backaction $(S_\mathrm F^\mathrm{qba})^{1/2} = (2\hbar \omega_\mathrm m m_\mathrm{eff} \mathit{\Gamma}_\mathrm{eff}^\mathrm{decay})^{1/2}=1.1\,$aN\,Hz$^{-1/2}$ is approaching $(S_\mathrm F^\mathrm{mode})^{1/2} = 4.3\,$aN\,Hz$^{-1/2}$ measured at $n_\mathrm p = 1.4\cdot 10^7$ for device A.
Force sensing with resonators based on two-dimensional materials hold promise for detecting electron and nuclear spins~\cite{rugar2004} using superconducting cavities compatible with relatively large magnetic fields~\cite{samkharadze2015}, and studying the thermodynamic properties of two-dimensional materials, such as the quantum capacitance and the magnetization~\cite{chen2015}.

\section{Methods}
\subsection{Cavity characterization}
In Figs.~\ref{fig2}d,e we plot the coefficient $|S_\mathrm{11}|^2$ and the phase of the reflected signal when sweeping the frequency over the cavity resonance at $\omega_\mathrm{c}/2\pi=7.416\,$GHz. To extract the external coupling rate $\kappa_\mathrm{ext}$ and the internal loss rate $\kappa_\mathrm{int}$ we fit the measurement with the line shape expected for a one-port reflection cavity \cite{aspelmeyer2014}\\
\begin{equation}
	S_{11}=\frac{\kappa_\mathrm{int}-\kappa_\mathrm{ext}-2i(\omega-\omega_\mathrm c)}{\kappa_\mathrm{int}+\kappa_\mathrm{ext}-2i(\omega-\omega_\mathrm c)},
	\label{eq:cavity}
\end{equation}\\
which yields $\kappa_\mathrm{int}/2\pi=950\,$kHz and
$\kappa_\mathrm{ext}/2\pi=850\,$kHz at $V_\mathrm g =3.002\,$V for device A. The rates of Device B are $\kappa_\mathrm{int}/2\pi=800\,$kHz and $\kappa_\mathrm{ext}/2\pi=1700\,$kHz at $V_\mathrm g = 0\,$V.

%\section{Data availability}
%The data that support the findings of this study are available from the corresponding author upon request.

\section{Acknowledgements}
We thank P.~Verlot and M.~Dykman for discussions. We acknowledge financial support by the ERC starting grant 279278 (CarbonNEMS), the EE Graphene Flagship (contact no. 604391), from MINECO and the Fondo Europeo de Desarrollo Regional (FEDER) through grant MAT2012-31338 and FIS2015-69831-P, the Fundaci\'o Privada Cellex, the Severo Ochoa Excellence Grant, and the Generalitat through AGAUR.

\section{Author contributions}
P.W. fabricated the devices, the process being developed by P.W. and J.G. P.W., J.G. and A.N. carried out the experiment with support from J.V.C. The data analysis was done by P.W. and J.G. with inputs from A.B. The experimental setup was built by J.G. with support from P.W. P.W. and A.B. wrote the manuscript with comments from J.G. and A.N. A.B. and J.G. conceived the experiment and supervised the work.

%\textbf{Supplementary information} accompanies this paper at %\hyperlink{http://www.nature.com/ncomms/index.html}{http://www.nature.com/ncomms}.\\

\textbf{Competing financial interests:} The authors declare no competing financial interests.

\end{document}